\documentclass{appolb}
\usepackage{graphicx}
\usepackage[utf8]{inputenc}
\usepackage{hyperref}
\usepackage{geometry}
\usepackage[T1]{fontenc}
\usepackage[english]{babel}
\usepackage{graphicx}
\usepackage{amsmath}
\usepackage{amssymb}
\usepackage{setspace}
\usepackage{fancyhdr}
\usepackage{wrapfig}
\usepackage{subfig}
\usepackage{hyperref}
\usepackage{sidecap}
\usepackage{theorem}
\usepackage{thc}
\usepackage{url}
\usepackage{booktabs}
\usepackage{multirow}
\usepackage{slashed}
\usepackage{longtable}

\begin{document}
\title{The flavour problem and family symmetry beyond the Standard Model%
\thanks{Presented  by M.Richter at the XXXVII International Conference of Theoretical Physics ``Matter to the Deepest 2015'', Ustroń, Poland, September 13-18,2015}%
}
\author{Bartosz Dziewit, Jacek Holeczek, Monika Richter, \\ Marek Zrałek
\address{Institute of Physics,University of Silesia\\
Uniwersytecka 4, 40-007 Katowice, Poland}
\\
{Sebastian Zając
}
\address{Faculty of Mathematics and Natural Studies,\\ Cardinal Stefan Wyszyński  University in Warsaw\\
Dewajtis 5, 01-815 Warsaw, Poland}
}

\maketitle
\begin{abstract}
In the frame of two Higgs doublet model we try to explain the lepton masses and mixing matrix elements assuming that neutrinos are Dirac particles. Discrete family symmetry groups, which are subgroups of U(3) up to the 1025 order are considered. Like in the one Higgs Standard Model, we found that discrete family symmetries do not give satisfactory answer for this basic questions in the flavour problem.
\end{abstract}
  
\section{Introduction}
The Standard Model is so far the best theory describing  particles and their interactions. However, it struggles to explain many problems.  Only in the lepton sector we have 10 arbitrary parameters (6 masses+3 mixing angles +1 CP phase) in the case of Dirac neutrinos and even 12 parameters (6 masses +3 mixing angles+3 CP phases) if we assume neutrinos to be of Majorana nature. Moreover, the  existence of 3 families of quarks and leptons, the nature of neutrinos, mechanism of neutrino mass generation and values of CP violating phases are still unresolved mysteries within the framework of this theory. The enumerated problems are the part of the so-called ``flavour problem'' \cite{fromth}. 

  Until 2012 it was thought that we were on the right path to find a solution: the TriBiMaximal (TBM) mixing \cite{Tbm} fully explained parameters determining the PMNS mixing matrix. However, thanks to more precise measurements \cite{reactor, Abe:2013hdq,Adamson:2013whj, Ahn:2012nd},it has been discovered that the reactor angle cannot be assumed to be zero. This fact resulted in the need to find another pattern  describing  mixing. After 2012, many ideas aimed at solving this problem, most of them were based on the simple extension of the Standard Model  by an addition of a discrete symmetry group \cite{lam,grimus, holt}. It was noticed that in order to get non-trivial mixing the family symmetry must be broken into two residual symmetries generating separately the forms of mass matrices in the charged  lepton and neutrino sector. In our opinion this idea is not so convincing: we impose full symmetry which next must be broken into its subgroups. Anyway, to our knowledge, that reasoning didn't lead to any reasonable results: the form of the mixing matrix wasn't clarified, any prediction for the masses hasn't been obtained.
  
  All these aspects motivated us to give a try a  different approach. Our idea consists in extending not only symmetry of the Standard Model but also the scalar sector. This concept has been already briefly described in \cite{dziewit}.
\section{The symmetry group}
As it has been already written in the introduction, our model focuses on examination of the consequences of adding  to  $SU(3)_{C}\times SU(2)_{L}\times U(1)_{Y}$  gauge symmetry  some extra group $G_{F}$  with some modification of the Higgs part of Lagrangian.
In our approach, we assume that $G_F$  (flavour symmetry) is finite and non-abelian. The reason for this choice is quite obvious: $TBM$ pattern was based on $A_{4}$ group. The long-lasting success of this model gave us some indications to search for solution in that ``area''.

  Apart from that, we expect from $G_{F}$ to be the subgroup of $U(3)$. Another condition for the symmetry comes from the  Yukawa Lagrangian:
\begin{equation}
\mathcal{L}_{Y}=-\sum_{\alpha,\beta=e,\mu,\tau}\sum_{j=1,2}[(h^{l}_j)_{\alpha\beta}\bar L_{\alpha L}\phi_{j}l_{\beta R}+(h^{\nu}_j)_{\alpha\beta } \bar L_{\alpha L}\tilde\phi_{j}\nu_{\beta R}]+H.c.,
\label{yukawa}
\end{equation}

where  $h^{l},h^{\nu}$ are Yukawa couplings, $L_{L}$ denotes  left-handed lepton doublet, $\nu_{R},l_{R}$ stand for right-handed singlets of $SU(2)_{L}\times U(1)_{Y}$ (of neutrino and charged lepton field respectively). Note that instead of one Higgs field we put here the sum over 2 doublets ($\tilde \phi=i\sigma_{2}\phi^{*}$). This is the extension of the scalar sector, which we mentioned before.

From the picture presented above the last requirement for the group naturally arises. We want the following fields'  transformation laws to be valid:
\begin{equation}
L'_{\alpha L}=(A^{L})_{\alpha,\chi} L_{\chi L},\quad l'_{\beta R}=(A^{l})_{\beta,\gamma}l_{\gamma R},\quad \nu'_{\beta R}=(A^{\nu})_{\beta,\delta}\nu_{\delta R}, \quad \phi'_{i}=(A^{\phi})_{i,k}\phi_{k},
\label{law}
\end{equation}
where $A^{L},A^{l}$,$A^{\nu}$,$A^{\phi}$ are irreducible representations of the flavour group $G_{F}$.
Thus, it becomes quite clear, that 2 and 3 dimensional irreducible representations of $G_{F}$ are needed. $A^{L}$, $A^{l}$ and $A^{\nu}$ should be 3-dimensional matrices, while $A_{\phi}$ must be 2-dimensional matrix with regard to the existence of two Higgs doublets.

At this point attributes of the wanted symmetry can be easily established. To sum up, the desired group  should be finite and non-abelian, be a subgroup of U(3) and possess 2 and 3-dimensional irreducible representations. At this stage, to simplify our search, we take into account only finite groups which possess at least one faithful 3-dimensional irreducible\footnote{There are also many finite subgroups of U(3) which possess a faithful 3-dimensional \textsl{reducible} representation but do not possess any faithful 3-dimensional \textsl{irreducible} representation.} representation and which cannot be written as a direct products with cyclic groups. In order to find the groups fulfilling all these requirements it is convenient to make use of GAP, which is the program for discrete algebra computation \cite{GAP}. Besides, since we are interested in finding the groups of small orders, the application of \textsl{the Small Group Library} \cite{small, small2} together with the $REPSN$\cite{repsn} package, which provides repressentations, seems to be necessary. Among all groups implemented in this library, up to the order 1025, we have found that 62 groups are of the kind we search for, among them 17 are subgroups of SU(3).

All considered subgroups of $U(3)$ can be identified with the classification of finite subgroups by Blichfeldt, Miller and Dickson \cite{su(3)} and Ludl \cite{ludl}.
The results are presented in the Table 1.
{\small
\begin{longtable}{|c|c|c|c|}
\hline
\textbf{[[i,j]]}&\textbf{The Group Description} &\textbf{The Group classification}& \textbf{ SU(3)?}\\
\hline
[[24,12]]&$ S_{4}$&$\Delta(24)=\Delta(6\times 2^{2})$ &\checkmark\\
\hline
[[48,30]]& $A_{4}\rtimes C_{4}$&$S_{4}(2)$& \\
\hline
[[54,8]]&$ (( C_{3}\times C_{3})\rtimes C_{3})\rtimes C_{2}$&$\Delta(54)=\Delta(6\times 3^{2})$ & \checkmark\\
\hline
[[96,64]] &$(( C_{4}\times C_{4})\rtimes C_{3})\rtimes C_{2}$&$\Delta(96)=\Delta(6\times 4^{2})$ & \checkmark\\
\hline
[[96,65]] & $A_{4}\rtimes C_{8}$&$S_{4}(3)$&  \\
\hline
[[108,11]]& $(( C_{3}\times C_{3})\rtimes C_{3})\rtimes C_{4}$&$\Delta(6\times 3^{2},2)$&  \\
\hline
[[150,5]]&$ (( C_{5}\times C_{5})\rtimes C_{3})\rtimes C_{2}$&$\Delta(150)=\Delta(6\times 5^{2})$ & \checkmark\\
\hline
[[162,10]]&$((( C_{3}\times C_{3})\rtimes C_{3}\times C_{3})\rtimes C_{3})\rtimes C_{2}$&  & \\
\hline
[[162,12]]&$(( C_{9}\times C_{3})\rtimes C_{3})\rtimes C_{2}$& & \\
\hline
[[162,14]]&$ (( C_{9}\times C_{3})\rtimes C_{3})\rtimes C_{2}$& $D(9,1,1;2,1,1) $& \checkmark\\
\hline
[[162,44]]&$(( C_{9}\times C_{3})\rtimes C_{3})\rtimes C_{2}$&$\Delta'(6\times 3^{2},2,1)$&  \\
\hline
[[192,182]]&$(( C_{4}\times C_{4})\rtimes C_{3})\rtimes C_{4}$&$\Delta(6\times 4^{2},2)$&  \\
\hline
[[192,186]] &$ A_{4}\rtimes C_{16}$&$ S_{4}(4)$&  \\
\hline
[[216,17]] &$(( C_{3}\times C_{3})\rtimes C_{3})\rtimes C_{8}$&$\Delta(6\times 3^{2},4)$&  \\
\hline
[[216,88]]&$ (( C_{3}\times C_{3})\rtimes C_{3})\rtimes Q_{8}$&$\Sigma(72\phi)$ &\checkmark\\
\hline
[[216,95]]&$(( C_{6}\times C_{6})\rtimes C_{3})\rtimes C_{2}$&$\Delta(216)=\Delta(6\times 6^{2})$  &\checkmark\\
\hline
[[294,7]]&$(( C_{7}\times C_{7})\rtimes C_{3})\rtimes C_{2}$&$\Delta(294)=\Delta(6\times 7^{2}) $ &\checkmark\\
\hline
[[300,13]]&$(( C_{5}\times C_{5})\rtimes C_{3})\rtimes C_{4}$&$\Delta(6\times 5^{2},2)$&  \\
\hline
[[324,13]]&$((( C_{3}\times C_{3})\rtimes C_{3})\rtimes C_{4})\rtimes C_{2}$& & \\
\hline
[[324,15]]&$(( C_{9}\times C_{3})\rtimes C_{3})\rtimes C_{4}$& &  \\
\hline
[[324,17]]&$((( C_{3}\times C_{3})\rtimes C_{3})\rtimes C_{4})\rtimes C_{2}$& &  \\
\hline
[[324,102]]&$(( C_{9}\times C_{3})\rtimes C_{3})\rtimes C_{4}$&$ \Delta'(6\times 3^{2},2,2)$&  \\
\hline
[[384,568]]&$(( C_{8}\times C_{8})\rtimes C_{3})\rtimes C_{2}$&$\Delta(384)=\Delta(6\times 8^{2}) $& \checkmark\\
\hline
[[384,571]]&$(( C_{4}\times C_{4})\rtimes C_{3})\rtimes C_{8}$&$\Delta(6\times 4^{2},3)$& \\
\hline
[[384,581]]&$A_{4}\rtimes C_{32}$&$S_{4}(5)$&  \\
\hline
[[432,33]]&$(( C_{3}\times C_{3})\rtimes C_{3})\rtimes C_{16}$& $\Delta(6\times 3^{2},4)$&  \\
\hline
[[432,239]]&$((( C_{3}\times C_{3})\rtimes C_{3})\rtimes C_{4})\rtimes C_{4}$& &  \\
\hline
[[432,260]]&$(( C_{6}\times C_{6})\rtimes C_{3})\rtimes C_{4}$&$\Delta(6\times 6^{2},2)$&  \\
\hline
[[486,26]]&$(( C_{27}\times C_{3})\rtimes C_{3})\rtimes C_{2}$& &  \\
\hline
[[486,28]]&$(( C_{27}\times C_{3})\rtimes C_{3})\rtimes C_{2}$& &  \\
\hline
[[486,61]]&$(( C_{9}\times C_{9})\rtimes C_{3})\rtimes C_{2}$&$\Delta(486)=\Delta(6\times 9^{2})$ & \checkmark\\
\hline
[[486,125]]&$(( C_{9}\times C_{3})\times C_{3})\rtimes C_{3})\rtimes C_{2}$& &  \\
\hline
[[486,164]]&$(( C_{27}\times C_{3})\rtimes C_{3})\rtimes C_{2}$&$\Delta'(6\times 3^{2},3,1)$&  \\
\hline
[[588,16]]&$(( C_{7}\times C_{7})\rtimes C_{3})\rtimes C_{4}$&$\Delta(6\times 7^{2},2)$ &\\
\hline
[[600,45]]& $(( C_{5}\times C_{5})\rtimes C_{3})\rtimes C_{8}$&$\Delta(6\times 5^{2},4)$ &\\
\hline
[600,179]]& $(( C_{10}\times C_{10})\rtimes C_{3})\rtimes C_{2}$ &$\Delta(600)=\Delta(6\times 10^{2})$ &\checkmark\\
\hline
[[648,19]]& $(((C_{3} \times C_{3}) \rtimes C_{3}) \times C_{8}) \rtimes C_{3}$ & & \\
\hline
[[648,21]]& $ (( C_{9}\times C_{3})\rtimes C_{3})\rtimes C_{8}$ & &\\
\hline
[[648,23]]& $(((C_{3} \times C_{3}) \rtimes C_{3}) \times C_{8}) \rtimes C_{3}$ & &\\
\hline
[[648,244]]& $(( C_{9}\times C_{3})\rtimes C_{3})\rtimes C_{8}$ &$\Delta'(6\times 3^{2},2,3)$ &\\
\hline
[[648,259]]& $ (( C_{18}\times C_{6})\rtimes C_{3})\rtimes C_{2}$ & $D(3,1,2;9,3,2)$&\checkmark\\
\hline
[[648,260]]& $(( C_{18}\times C_{6})\rtimes C_{3})\rtimes C_{2}$ & &\\
\hline
[[648,266]]& $  ((C_{6} \times C_{6} \times C_{3}) \rtimes C_{3}) \rtimes C_{2}$ & &\\
\hline
[[648,531]]&$ C_{3} . (((C_{3} \times C_{3})\rtimes Q_{3}) \rtimes C_{3}) $& & \\
\hline
[[648,532]] &$(((C_{3} \times C_{3}) \rtimes C_{3}) \times Q_{8}) \rtimes C_{3} $&$\Sigma(216\phi)$ &\checkmark \\
\hline
[[648,533]]& $(((C_{3} \times C_{3}) \rtimes C_{3}) \times Q_{8}) \rtimes C_{3} $& & \\
\hline
[[648,551]]& $((C_{9} \times C_{3}) \rtimes C_{3}) \rtimes Q_{8}$& & \\
\hline
[[648,563]]& $(( C_{18}\times C_{6})\rtimes C_{3})\rtimes C_{2}$ & & \\
\hline
[[726,5]]&$ ((C_{11} \times C_{11}) \rtimes C_{3}) \rtimes C_{2}$ &$\Delta(726)=\Delta(6\times 11^{2})$ &\checkmark \\
\hline
[[768,1085333]]& $(( C_{4}\times C_{4})\rtimes C_{3})\rtimes C_{16}$ &$\Delta(6\times 4^{2},8)$ & \\
\hline
[[768,1085335]]&$(( C_{8}\times C_{8})\rtimes C_{3})\rtimes C_{4}$ &$\Delta(6\times 8^{2},2)$ &\\
\hline
[[768,1085351]]& $ A_{4}\rtimes C_{64}$&$S_{4}(6)$ & \\
\hline
[[864,69]& $  (( C_{3}\times C_{3})\rtimes C_{3})\rtimes C_{32}$ &$\Delta(6\times 3^{2},16)$ & \\
\hline
[[864,675]]& $ (((C_{3} \times  C_{3}) \rtimes C_{3}) \rtimes C_{4}) \rtimes C_{8}$& & \\
\hline
[[864,701]]& $(( C_{12}\times C_{12})\rtimes C_{3})\rtimes C_{2}$&$\Delta(864)=\Delta(6\times 12^{2})$ &\checkmark \\
\hline
[[864,703]]& $(( C_{6}\times C_{6})\rtimes C_{3})\rtimes C_{8}$ &$\Delta(6\times 6^{2},4) $& \\
\hline
[[972,29]]& $(( C_{27}\times C_{3})\rtimes C_{3})\rtimes C_{4}$ & & \\
\hline
[[972,31]]& $(( C_{27}\times C_{3})\rtimes C_{3})\rtimes C_{4}$ & & \\
\hline
[[972,64]]&  $(( C_{9}\times C_{9})\rtimes C_{3})\rtimes C_{4}$ &$\Delta(6\times 9^{2},2)$ & \\
\hline
[[972,309]]& $ (((C_{9} \times C_{3}) \rtimes C_{3}) \rtimes C_{4}) \rtimes C_{3}$ & & \\
\hline
[[972,348]]& $ (( C_{27}\times C_{3})\rtimes C_{3})\rtimes C_{4}$ &$\Delta'(6\times 3^{2},3,2)$ & \\
\hline
[[1014,7]] & $((C_{13} \times C_{13}) \rtimes C_{3}) \rtimes C_{2}$ &$\Delta(1014)=\Delta(6\times 13^{2})$ &\checkmark \\
\hline
\caption{Classification of  considered subgroups of $U(3)$}
\end{longtable}
}
\section{Invariance equation and its interpretation}
At this moment, one can introduce the main concept, which is the requirement imposed on the  Yukawa Lagrangian (Eq.\eqref{yukawa}) to be invariant under transformations Eq.\eqref{law}. In other words, we demand the following relations to be fulfilled:
%
\begin{equation}
\sum_{i=1}^{2}(A^{\phi})_{ik}(A^{L\dag})_{\alpha\gamma}(h_{i}^{l})_{\gamma\delta}(A^{l})_{\delta\beta}=[h_{k}^{l}]_{\alpha\beta},\qquad \sum_{i=1}^{2}(A^{\phi})^{*}_{ik}(A^{L\dag})_{\alpha\gamma}(h_{i}^{\nu})_{\gamma\delta}(A^{l})_{\delta\beta}=[h_{k}^{\nu}]_{\alpha\beta}.
\label{pierwsze}
\end{equation}
We can easily simplify these relations  by swapping the indices:
\begin{equation}
\sum_{i=1}^{2} (A^{\phi})^{T}_{ki}(A^{L\dag})_{\alpha\gamma}(A^{l})^{T}_{\beta\delta}(h_{i}^{l})_{\gamma\delta}=[h_{k}^{l}]_{\alpha\beta},\qquad \sum_{i=1}^{2} (A^{\phi})^{\dag}_{ki}(A^{L\dag})_{\alpha\gamma}(A^{\nu})^{T}_{\beta\delta}(h_{i}^{\nu})_{\gamma\delta}=[h_{k}^{\nu}]_{\alpha\beta}.
\label{drugie}
\end{equation}
Then, it is possible to present expressions derived  in Eq.\eqref{drugie} as two eigenproblems :
\begin{equation}
\mathbf{N_{1}}\Gamma^{l}=\Gamma^{l},\qquad \mathbf{N_{2}} \Gamma^{\nu}=\Gamma^{\nu},
\label{eigen}
\end{equation}
where: 
\begin{equation}
\Gamma^l=\left( \begin{array}{c}
h^l_1 \\ 
h^l_2 \\ 
\end{array}
\right), \quad 
\Gamma^\nu=\left( \begin{array}{c}
h^\nu_1 \\ 
h^\nu_2 \\ 
\end{array}
\right),
\end{equation} 
and:
\begin{equation}
\mathbf{N_{1}}=(A^{\phi})^{T}\otimes (A^{L})^{\dag}\otimes (A^{l})^{T},\qquad \mathbf{N_{2}}=(A^{\phi})^{\dag}\otimes (A^{L})^{\dag}\otimes (A^{\nu})^{T}.
\end{equation}
Note that $\Gamma^l$, $\Gamma^\nu$ are the vectors composed of appropriate elements of matrices $h_k^l$ and $h_k^\nu$ respectively. In both cases  $N_{1}$ and $N_{2}$ are 18-dimensional. 

The way of solving  equations  Eq.\eqref{eigen} is described in detail in \cite{ludl2}. In general, the algorithm can be summarized as follows: construction of  $\mathbf{N_{1}}$ and $\mathbf{N_{2}}$ for all generators of the considered flavour group $G_{F}$, looking for the eigensubspace for all generators, determining the common eigensubspace, establishing the base vector of the  common eigensubspace.

To conclude, the base vector of the common eigensubspace constitute  the solution. However,  the first step introduced in the presented algorithm requires some comment.  It is only necessary to take the generators' representations into account during the calculations. This non-trivial fact results from the following proposition (see e.g. \cite{ludl2}): if the invariance equations Eq.\eqref{drugie} are valid for the representations of some generators $g_{1}$ and $g_{2}$ of the flavour group $G_{F}$, then they are automatically satisfied by the representations of their product $g_{3}=g_{1}g_{2}$.

It turns out that solution of the invariance equation has got simple mathematical interpretation. It can be proven (see e.g. \cite{ludl2}) that Yukawa couplings $h^{\nu}$ and $h^{l}$ play the role of Clebsch-Gordan coefficients for the following decompositions:
\begin{equation}
A^{L}\otimes (A^{l})^{*}=\oplus_{\lambda}A_{\lambda},\qquad A^{L}\otimes (A^{\nu})^{*}=\oplus_{\lambda} A_{\lambda},
\label{decomp}
\end{equation}
where in the first case we look for the 2-dimensional representation ($A^{\phi}=A_{\lambda}$ for some $\lambda$), while in the second case we demand the existence of $(A^{\phi})^{*}=A_{\lambda}$ for some $\lambda$. This condition is necessary to get some solution for the invariance equation.
\section{The mass matrices and mixing matrix in the lepton sector}
After finding appropriate Yukawa couplings: $h_{1,2}^{l}$ and $h_{1,2}^{\nu}$ one has got  everything to find out how mass matrices:
\begin{equation}
M^{l}_{\alpha,\beta}=\frac{1}{\sqrt{2}}\sum_{i=1}^{2} v_{i} (h^{l}_{i})_{\alpha,\beta},\quad M^{\nu}_{\alpha,\beta}=\frac{1}{\sqrt{2}}\sum_{i=1}^{2} v_{i} (h^{\nu}_{i})_{\alpha,\beta},
\label{mass}
\end{equation}
look like. Here $v_{i}$ stand for vacuum expectation values of the Higgs fields.

To proceed with the calculations one should diagonalize created matrices. It is commonly known in the literature that in order to derive the mass eigenvalues it is necessary to use biunitary transformation:
\begin{equation}
V_{L}^{l\dag}M^{l}V^{l}_{R}=M^{l}_{diag},\quad V_{L}^{\nu\dag}M^{\nu}V^{\nu}_{R}=M^{\nu}_{diag}.
\end{equation}

 In practice, one diagonalizes $M^{l}M^{l\dag}$ and $M^{\nu}M^{\nu\dag}$ instead of single $M^{\nu}$ or $M^{l}$ since as opposed to the latter case one needs  only one unitary matrix to perform the diagonalization:
$$
V_{L}^{l\dag}(M^{l}M^{l\dag})V_{L}^{l}=(M^{l}M^{l\dag})_{diag},\quad V_{L}^{\nu\dag}(M^{\nu}M^{\nu\dag})V_{L}^{\nu}=(M^{\nu}M^{\nu\dag})_{diag}.
$$

From the obtained matrices $V^{l}_{L}$ and $V^{\nu}_{L}$ one can easily create the mixing matrix:
\begin{equation}
U_{PMNS}=V_{L}^{l\dag}V_{L}^{\nu}.
\end{equation}

Now, we can try to find such a free model parameters, which give correct values of the lepton masses and mixing matrix elements \cite{data}.


The obtained preliminary results indicate that in the set of examined groups there are no such, which are able to reproduce the mixing matrix elements and lepton masses. 
\section{Conclusions}
Clearly, the results obtained so far aren't in agreement with the experiment.
 If it turned out that results match the experimental data
it would be  necessary  to have a second look at the Higgs potential. 
We would need to know whether the Higgs potential meets all experimental requirements, gives two different vacuum expectation values, and all additional Higgs particles satisfy existing experimental bounds: e.g. the flavour changing neutral current is small.

 
In the nearest future, we plan to study the model with another extensions of the scalar sector (for example with three Higgs doublets) and similar models for Majorana neutrinos. Apart from that, we are going to examine the model with left-right symmetry where additional sterile neutrinos are present.

This work has been supported by the Polish Ministry of Science and Higher Education under grant No. UMO-2013/09/B/ST2/03382.

\end{document}